\documentclass[11pt,twoside]{article}
\usepackage{graphicx,epsfig,natbib,epstopdf}
\usepackage{CS18}
\begin{document}
%
%
%
\title{Understanding Astrophysical Noise from Stellar Surface Magneto-Convection}
%
%
\author{H.\,M.\ Cegla$^{1}$, C.\,A.\ Watson$^{1}$, S.\ Shelyag$^{2}$,  M.\ Mathioudakis$^{1}$}
\affil{$^1$Astrophysics Research Centre, School of Mathematics \& Physics, Queen's University Belfast, University Road, Belfast BT7 1NN, UK}
\affil{$^2$Monash Centre for Astrophysics, School of  Mathematical Sciences, Monash University, Clayton, Victoria,  3800, AU}

%
\begin{abstract}

To obtain cm s$^{-1}$ precision, stellar surface magneto-convection must be disentangled from observed radial velocities (RVs). In order to understand and remove the convective signature, we create Sun-as-a-star model observations based off a 3D magnetohydrodynamic solar simulation. From these Sun-as-a-star model observations, we find several line characteristics are correlated with the induced RV shifts. The aim of this campaign is to feed directly into future high precision RV studies, such as the search for habitable, rocky worlds, with forthcoming spectrographs such as ESPRESSO.
\end{abstract}
\section{Introduction}
Astrophysical noise introduces spurious spectroscopic signals that hinder the achievable radial velocity (RV) precision. Such noise sources include starspots, plage/faculae, and magneto-convection, as well as oscillations, meridional flows, and potentially variable gravitational redshift \citep{saar97, schrijver00, beckers07, cegla12}. Even magnetically quiet stars, which may lack the large amplitude signals induced by starspots, still exhibit inhomogeneities due to stellar surface magneto-convection, as long as they have an outer convective envelope. In the convective envelope, hot bubbles of bright plasma (known as granules) rise to the stellar surface where they eventually cool, darken and sink down into the intergranular lanes. As the plasma bubbles rise through the stellar photosphere, and therefore move towards the observer, they are blueshifted and as they disperse and fall downwards into the stellar interior they are redshifted with respect to the observer. Since the granules are brighter and cover more surface area than the intergranular lanes the result is a net convective blueshift, which introduces asymmetries into the observed stellar lines (the result is a `C'-shaped absorption line bisector for solar-type stars). The exact ratio of granules to intergranular lanes is constantly changing as the granules evolve and interact with their surroundings, which in turn means the observed line asymmetries are constantly changing. For Sun-like stars, the net result from these induced line changes is RV shifts at the level of several 10s of cm s$^{-1}$ \citep{schrijver00}. It is clear that an understanding of convection on solar-type stars, and indeed for all low-mass stars with convective envelopes, is essential if we are to reach cm s$^{-1}$ RV precision. In these proceedings, we report on our study of the effects of stellar photospheric magneto-convection using state-of-the-art 3D magnetohydrodynamical (MHD) solar simulations and detailed spectroscopic diagnostics. Our goal is to disentangle and remove the convective induced RVs from stellar absorption lines.

\section{The Simulations}
\begin{figure}
\begin{center}
\includegraphics[scale=0.35, trim= 0.25cm 0cm 0.9cm 0cm, clip]{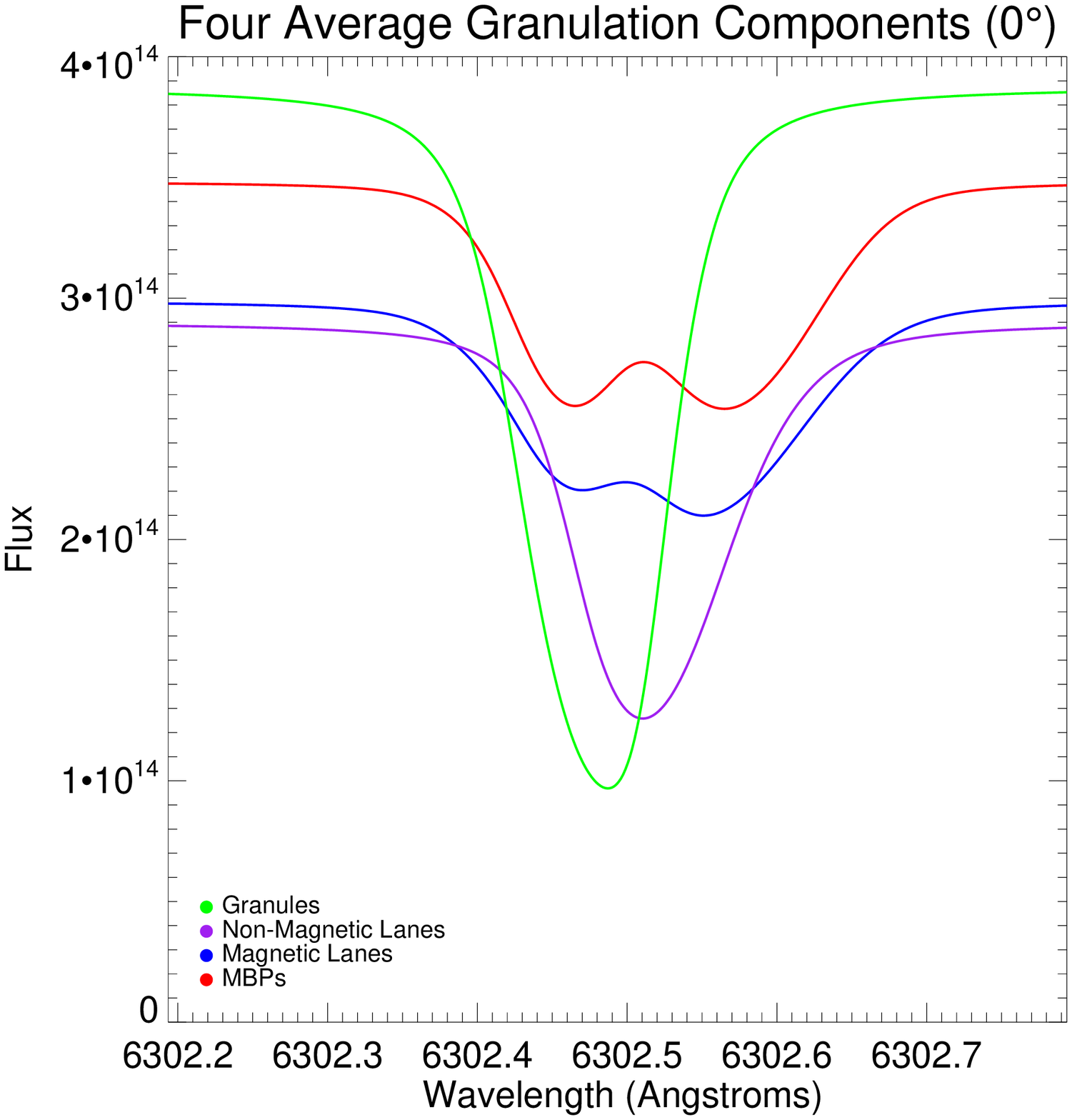}
\includegraphics[scale=0.35, trim= 0.25cm 0cm 0.9cm 0cm, clip]{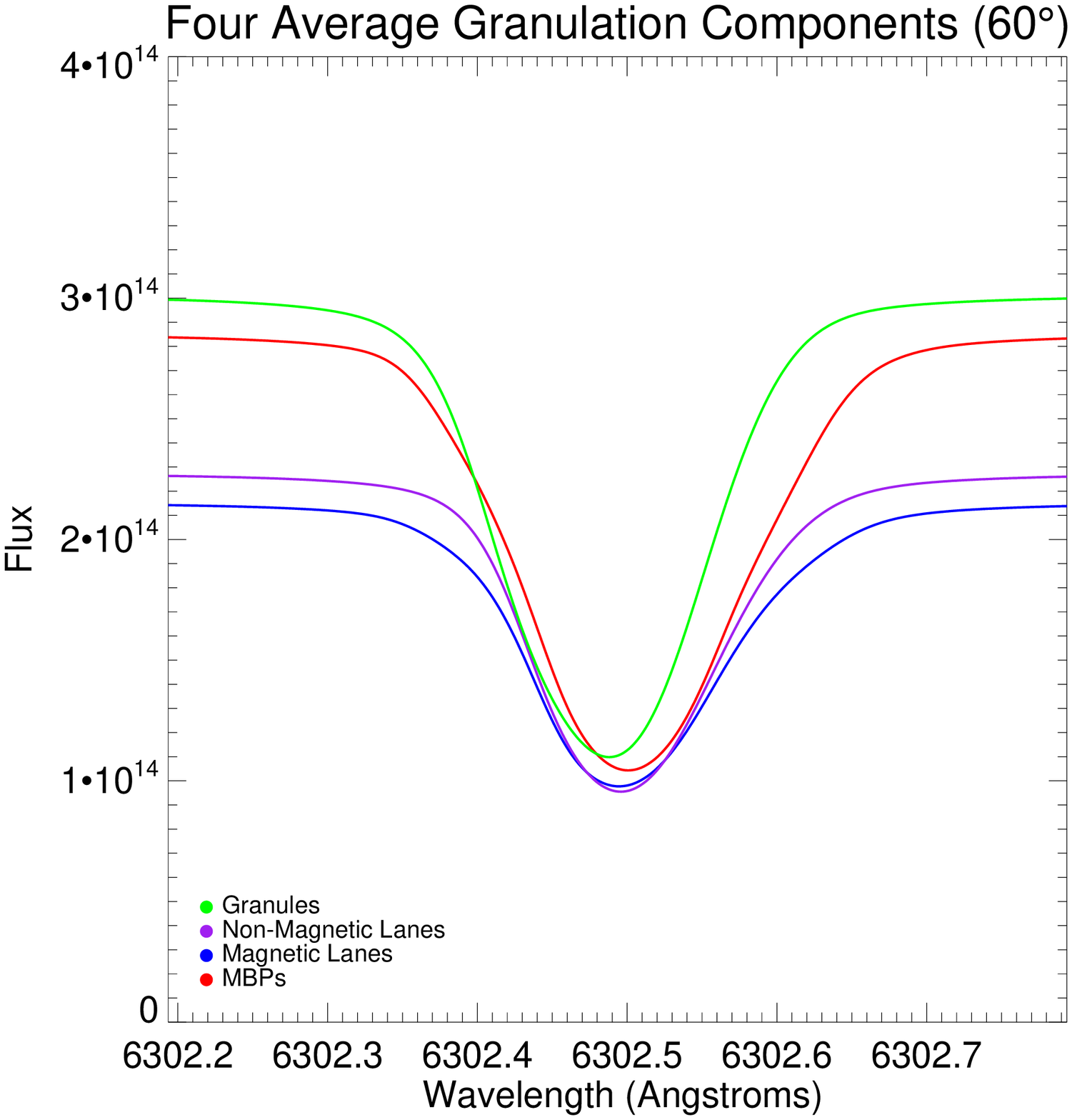}
\caption{Average line profiles from the four different contributions to granulation used in the parameterisation: granules (green), magnetic bright points (red), magnetic (blue) and non-magnetic (purple) intergranular lanes, for two different stellar disk center-to-limb angles: 0$\deg$ (left) and 60$\deg$ (right). \label{fig:4comp_l}}
\end{center}
\end{figure}

\begin{figure}
\begin{center}
\includegraphics[scale=0.4, trim= 0.25cm 0cm 0.4cm 0cm, clip]{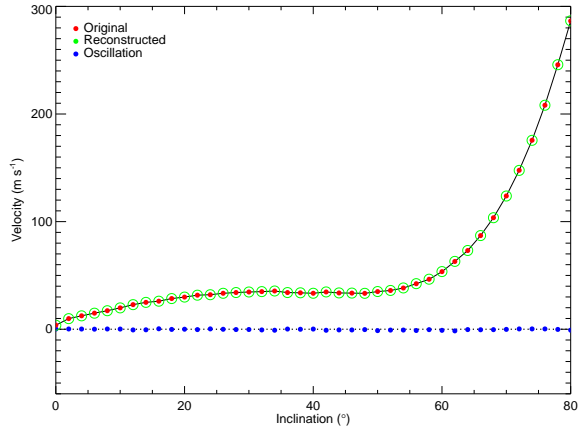}
\caption{The average granulation RVs over the time-series, for both the original profiles from the simulations (red points, with a black solid line connecting them) and the reconstructed profiles (green circles), relative to disk center and as a function of inclination. Also plotted are the average oscillation RVs (blue points), which are a natural byproduct of the simulation and were removed from the original profiles to obtain the simulated granulation RVs. The oscillation RVs are near zero because averaging over the time-series removes much of the oscillation signal. The black dotted line is a straight line at 0 m s$^{-1}$ to guide the eye. \label{fig:RV0dgr_l2}}
\end{center}
\end{figure}

We simulate the solar photosphere using the MURaM code \citep{voegler1} with a simulation box measuring 12 $\times$ 12 Mm$^2$ in the horizontal direction and 1.4 Mm in the vertical direction. Stokes-$I$ profiles for the $6302~\mathrm{\AA}$ FeI absorption line were calculated for 190 simulated photospheric model snapshots using the SPINOR code \citep{frutiger,shelyag07}. Due to computational constraints, and a desire to breakdown the underlying physics, we parameterise the granulation signal using an $\sim$ 80 minute time-series from the MHD simulation \citep{cegla13}. 
This is done by separating the pixels, and their corresponding Stokes-$I$ Fe I 6302 $\AA$ absorption line profiles, within each snapshot based on continuum intensity and magnetic field limits. The result is four granulation component line profiles (representing granules, non-magnetic intergranular lanes, magnetic intergranular lanes, and magnetic bright points) that can be used to generate Stokes-$I$ profiles with convection induced asymmetries as long as the appropriate filling factors are used (see Figure~\ref{fig:4comp_l}). Since granulation is naturally corrugated, observations at different line-of-sight (LOS) angles view different aspects of the granulation. As one views granulation toward the stellar limb the granular walls become visible and some of the intergranular lanes are obstructed by the granules. Furthermore, once horizontal flows at disk center now have components along the LOS and hence the four component profiles experience both line shape changes and relative velocity shifts. To incorporate these variations, we incline the MHD snapshots from 0-80$\deg$ in 2$\deg$ steps and repeat the parameterisation at each inclination step (computing the simulations beyond 80$\deg$ was computationally unfeasible). An example of the variations in the four component profile shapes is shown on the right side of Figure~\ref{fig:4comp_l}, whereas the RVs induced across the stellar limb due to the corrugated nature of granulation is shown in Figure~\ref{fig:RV0dgr_l2}. This RV dependency on the convective blueshift term as a function of limb angle could be important for current methods of removing activity signals.

To ensure that our parameterization is representative of the MHD simulations we use the filling factors from the aforementioned time-series in conjunction with the four component profiles to reconstruct the average absorption line profiles from the simulation snapshots. The average relative error is measured to ensure the line shape of the reconstructed profiles matches the original profiles from the simulation. The velocity shifts due to granulation are also measured in each instance (\citet{cegla13} provides more details on this process for disk center -- note this process remains the same for each inclination across the disk). For all inclinations the reconstructed profiles agree with the original profiles to better than 0.8\%. The average RV residual (the difference between granulation RVs from the original simulation and those from the reconstruction) only varies from $\sim$ 3 to -5 cm s$^{-1}$ on a 100 to 20 m s$^{-1}$ signal. As a result, we are confident that our granulation parameterisation is representative of the more computationally expensive MHD simulations. 

\section{Sun-as-a-star Model Observations}
\label{sec:sunasstar}
We used the granulation component filling factor distributions and relationships determined from the MHD simulations to generate new line profiles that are representative of those produced by the simulation itself. To begin, we randomly selected a granule filling factor based on the cumulative distribution function (CDF) from the original time-series. From the MHD simulations we found a strong linear relationship between granules and non-magnetic intergranular lanes, and thus we used this relationship to then select the non-magnetic intergranular lane filling factor. Unfortunately, there were no strong correlations between these two components and the magnetic components so the magnetic bright point filling factors were selected at random based on their CDF from the original time-series. Finally, the fourth component (magnetic intergranular lanes) was determined by the fact that all four component filling factors  must sum to one. Kolmogorov-Smirnov (K-S) and Wilcoxon Rank-Sum tests were performed to ensure that the generated filling factors belong to the same population as that produced from the MHD simulation (all components passed at all stellar limb angles, except the magnetic intergranular lanes at $>$ 20$\deg$, but this component contributes very little to the total profiles at these angles and thus this effect is likely negligible). 

To simulate a Sun-as-a-star model observation, we tiled a stellar disk with 12 $\times$ 12 Mm$^2$ regions, with the appropriate projected area, and randomly assigned each tile a newly generated line profile. Note that stellar limb darkening was accounted for within the original simulations and therefore also in the four component line profiles; however, we did linearly interpolate between stellar limb angles to correct for the odd inclinations and those inclinations greater than 80$\deg$ (due to computational constraints, tiles between 80-90$\deg$ were all assigned a line profile based on the four components at 80$\deg$ -- note this effect is likely negligible due to the small projected area and heavy limb darkening in this region). Once all the tiles were assigned an appropriate line profile we then integrated over the entire disk to simulate a model observation. This process was repeated 1,000 times to build up a statistically significant sample size to study. Each observation is independent and therefore relates to the physical case where each observation was separated by at least one convective turnover timescale. In future work we aim to incorporate the effects of a non-zero exposure time. The disk-integrated line profiles and their line bisectors are shown in the left side of Figure~\ref{fig:diskint_Vasy}. 

\section{Analysis and Results}
Finally, we examined the Sun-as-a-star disk-integrated line profiles for any correlation between line shape change or asymmetries and convection induced RV. To begin, we inspected numerous characteristics of the line bisector (including bisector inverse slope, bisector curvature, velocity displacement, and bisector amplitude [\citet{queloz01, dall06,  povich01}]), as well as line depth and line width at various depths. We also investigated diagnostics that addressed the entirety of the line profile, such as bi-Gaussian fitting and V$_{asy}$ \citep{fig13}. Finally, we integrated the area under the line profile to create a proxy for photometric brightness. 

\begin{center}
\begin{figure}[b!]
\includegraphics[scale=0.4, trim= 1.1cm 0cm 1.5cm 0cm, clip]{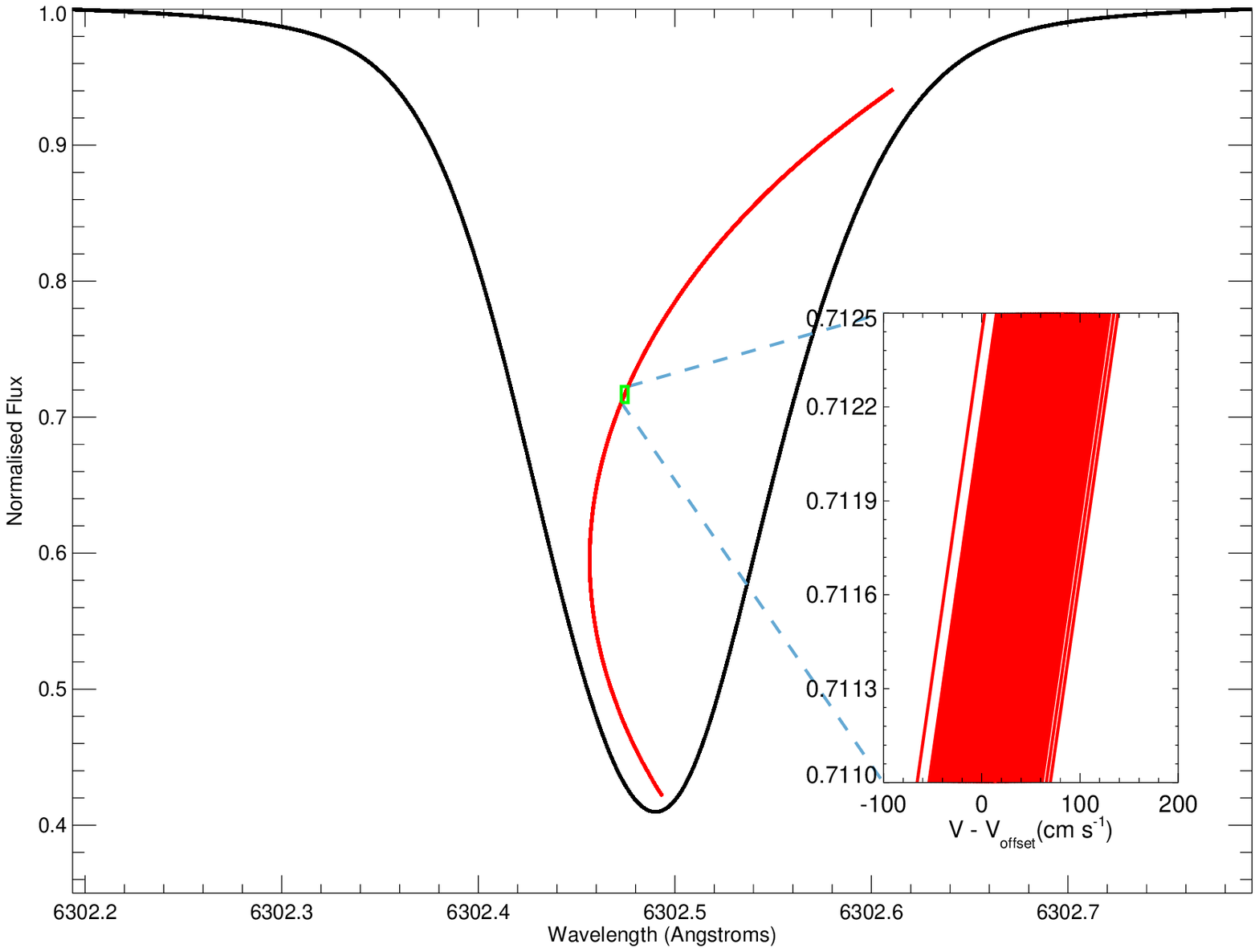}
\includegraphics[scale=0.37, trim= 1.3cm 0cm 0.9cm 0.5cm, clip]{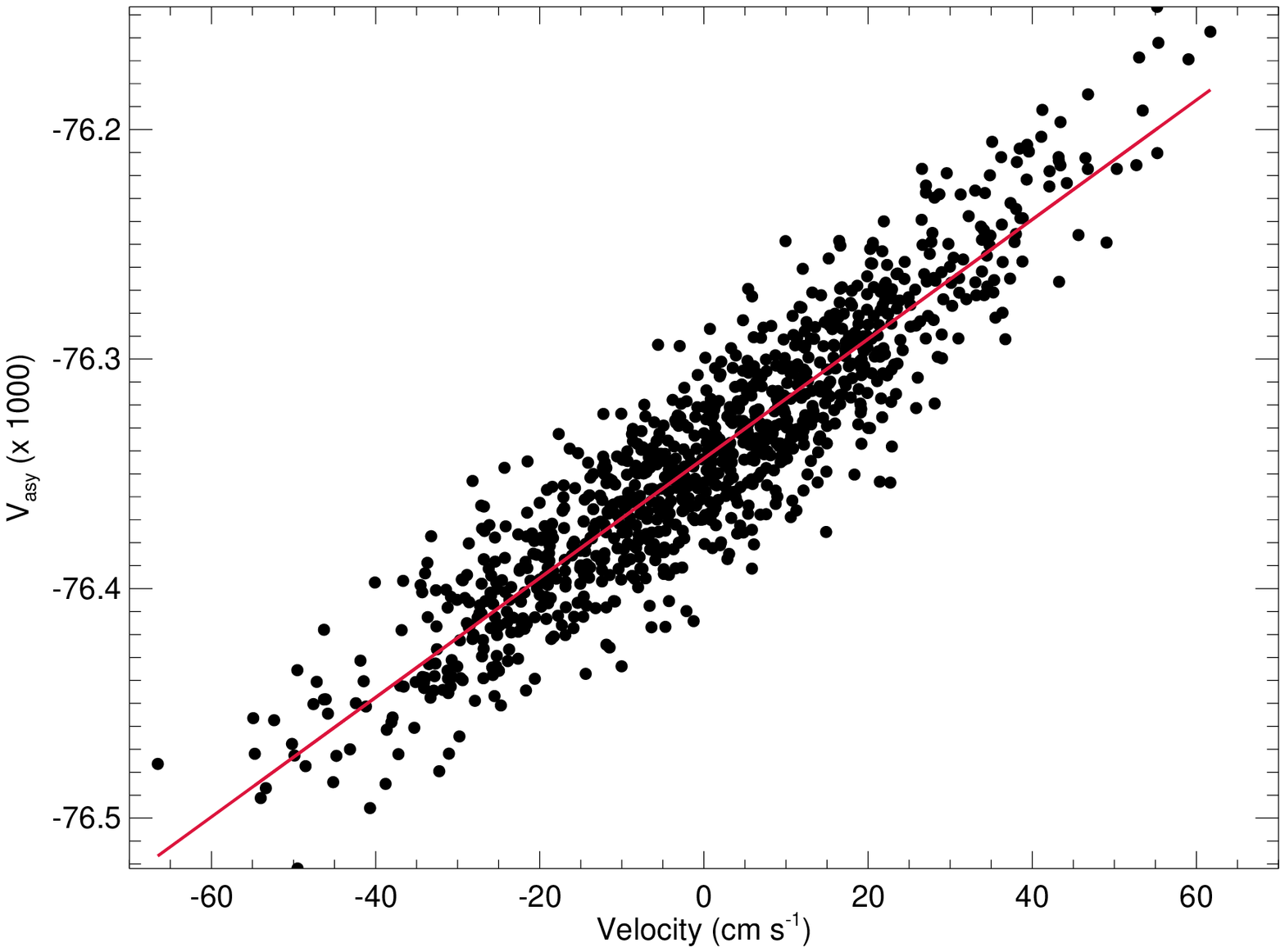}
\caption{Left: Each of the 1,000 disk-integrated line profiles from the photospheric magneto-convective model star observations (with an average magnetic field of 200 G) have been overplotted with their corresponding line bisectors (in red). The full line bisectors have been amplified by a factor of 30 so that the C-like curvature is visible. The profiles/bisectors are so similar that only 1 profile/bisector is discernible. In order to view the changes, the inset further amplifies, by a factor of $\sim$ 10,000, a small section of the line bisectors (marked by a green rectangle) where the Doppler shift (V) of the bisector positions has been calculated relative to the central wavelength; the velocity of the first profile (V$_{offset}$) has been subtracted from all the profiles to show the relative velocity variation. The variation of $\pm$ $\sim$ 50 cm s$^{-1}$ seen here is inline with predictions for solar granulation induced RV variability \citep[e.g.][]{meunier10, dumusque11}. Right: V$_{asy}$ vs RV for the 1,000 disk-integrated granulation line profiles. The solid red line indicates a robust linear regression fit. The Pearson's R correlation coefficient is $\sim$ 0.91. \label{fig:diskint_Vasy}}
\end{figure}
\end{center}

V$_{asy}$ and the proxy for photometric brightness produced the strongest correlations with RV and largest noise reduction (determined by subtracting off the correlation and comparing the standard deviation of the RVs before and after), a $\sim$ 50\% noise reduction in total. A 50\% reduction in convective noise meant the overall noise level was around 10 cm s$^{-1}$ in these quiet star simulations. The right side of Figure~\ref{fig:diskint_Vasy} shows the excellent correlation between V$_{asy}$ and granulation induced RV. The next best correlations were with the line depth and bisector curvature, which are able to remove $\sim$ 35\% of the noise. The remaining diagnostics either remove less than 20\% of the noise (velocity displacement, bisector amplitude, equivalent width) or actually result in an increase of the overall noise (bisector inverse slope and bi-Gaussian fitting). 

These initial results indicate hope that we will be able to empirically measure the convective induced line shape variations and use them to remove the convective induced RVs in high precision studies. It is important to note that we are currently working on an updated version of the radiative MHD solar simulation (with updated equation of state and opacities etc.) to try to match the disk integrated line profile more closely with solar observations. Initial results from the updated MHD simulation indicate there is a slight change in bisector shape and in some of the diagnostics, but strong correlations are still present and we anticipate similar overall results in noise reduction. We are also working to incorporate lower magnetic field strengths, additional sources of noise, finite exposure lengths, and the effects of planets. Additionally, we plan to extend the simulations across a suite of stellar lines (with a wide range of formation heights, absorption strengths, excitation and ionisation potentials) and various spectral types (M-F). Simultaneously, we are also working to test these results observationally with the Sun and with the highest precision stellar targets available with the overall aim to remove magneto-convective noise to a sub 10 cm s$^{-1}$ level in time for the ESPRESSO spectrograph to come online. 

\acknowledgments{
HMC acknowledges support from the Leverhulme Trust grant RPG-249. CAW would like to acknowledge support by STFC grant ST/I001123/1. This research was undertaken with the assistance of resources provided at the NCI National Facility systems at the Australian National University, supported by Astronomy Australia Limited, and at the Multi-modal Australian ScienceS Imaging and Visualisation Environment (MASSIVE) (www.massive.org.au). SS gratefully thanks the Centre for Astrophysics \& Supercomputing of Swinburne University of Technology (Australia) for the computational resources provided. SS is the recipient of an Australian Research Councils Future Fellowship (project number FT120100057).
}


\normalsize

\begin{references}

\bibitem[Cegla et al.(2012)]{cegla12} Cegla, H.~M., Watson, 
C.~A., Marsh, T.~R., et al.\ 2012, \mnras, 421, L54 

\bibitem[Cegla et al.(2013)]{cegla13} Cegla, H.~M., Shelyag, 
S., Watson, C.~A., \& Mathioudakis, M.\ 2013, \apj, 763, 95 

\bibitem[Beckers(2007)]{beckers07} Beckers, J.~M.\ 2007, 
Astronomische Nachrichten, 328, 1084 

\bibitem[Dall et 
al.(2006)]{dall06} Dall, T.~H., Santos, N.~C., Arentoft, T., Bedding, T.~R., \& Kjeldsen, H.\ 2006, \aap, 454, 341

\bibitem[Dumusque et 
al.(2011)]{dumusque11} Dumusque, X., Lovis, C., S{\'e}gransan, D., et al.\ 2011, \aap, 535, A55 

\bibitem[Figueira et 
al.(2013)]{fig13} Figueira, P., Santos, N.~C., Pepe, F., Lovis, C., \& Nardetto, N.\ 2013, \aap, 557, A93 

\bibitem[Frutiger et al.(2000)]{frutiger} Frutiger, C., Solanki, S.~K., Fligge, M., \& Bruls, J.~H.~M.~J.\ 2000, \aap, 358, 1109 

\bibitem[Meunier et 
al.(2010)]{meunier10} Meunier, N., Desort, M., \& Lagrange, A.~M.\ 2010, \aap, 512, A39 

\bibitem[Povich et al.(2001)]{povich01} Povich, M.~S., Giampapa, 
M.~S., Valenti, J.~A., et al.\ 2001, \aj, 121, 1136 

\bibitem[Queloz et 
al.(2001)]{queloz01} Queloz, D., Henry, G.~W., Sivan, J.~P., et al.\ 2001, \aap, 379, 279 

\bibitem[Saar 
\& Donahue(1997)]{saar97} Saar, S.~H., \& Donahue, R.~A.\ 1997, \apj, 485, 319 

\bibitem[Schrijver 
\& Zwaan(2000)]{schrijver00} Schrijver, C.~J., \& Zwaan, C.\ 2000, Solar and stellar magnetic activity (New York: Cambridge University Press), 34

\bibitem[Shelyag et 
al.(2007)]{shelyag07} Shelyag, S., Sch{\"u}ssler, M., Solanki, S.~K., V{\"o}gler, A.\ 2007, \aap, 469, 731

\bibitem[V{\"o}gler et al.(2005)]{voegler1} V{\"o}gler, A., Shelyag, S., Sch{\"u}ssler, M., et al.\ 2005, \aap, 429, 335 

\end{references}
\end{document}